\documentclass[review]{elsarticle}

\usepackage{lineno,hyperref}
\modulolinenumbers[5]

\journal{International Journal of Heat and Mass Transfer}

\usepackage{color}
\usepackage{multirow}
\usepackage{dcolumn}
\usepackage{multicol}
\usepackage{graphicx}
\usepackage{caption}
\usepackage{amssymb}
\usepackage{epstopdf}
\usepackage{mathtools}
\usepackage{float}
\usepackage{bm}

\biboptions{sort&compress}

\usepackage{color}

\def\exp{{\rm {exp}}}

\newcommand{\ks}{\textcolor{black}} 

\begin{document}

\begin{frontmatter}


\title{\ks{Modeling binary collision of evaporating drops}}
\author{Ashwani Kumar Pal$^a$, Kirti Chandra Sahu$^b$ and Gautam Biswas$^{a}$\footnote{gtm@iitk.ac.in} }
\address{$^a$Department of Mechanical Engineering, Indian Institute of Technology Kanpur, Kanpur - 208016, Uttar Pradesh, India  \\ 
$^b$Department of Chemical Engineering, Indian Institute of Technology Hyderabad, Sangareddy 502 284, Telangana, India}

\begin{abstract}
We investigate the interactions between two drops in a heated environment and analyze the effect of evaporation on bouncing, coalescence and reflexive separation phenomena. A reliable mass transfer model is incorporated in a coupled level-set and volume-of-fluid framework to accurately model the evaporation process and the evolution of drop interfaces during the interactions. The numerical technique is extensively validated against the benchmark problems involving the evaporation of a single drop. We analyze the contour plots of temperature and vapor mass fraction fields for each collision outcome. Our numerical simulations reveal that vapor entrapment during the separation process, with high-velocity vapor manages to escape. Increasing evaporation rates result in slower post-collision drop separation. Furthermore, the differences in kinetic energy and surface energy are analyzed for different Stefan numbers. The coalescence of drops exhibits energy oscillations until dissipation, while the bouncing and reflexive separations lack such oscillations. \ks{In the reflexive separation regime, the kinetic energy of the drops becomes zero after detachment.} 
\end{abstract}

\end{frontmatter}


\section{Introduction}\label{sec:introduction}

The collision of two or more drops is relevant in many industrial applications and environmental situations. The common scenarios can be observed in spray combustion \cite{chiu2000spray,smith1972}, spray drying \cite{zamani2021drying}, and rainfall \cite{rgun1965rain}. The outcome of the binary collision of drops may be different depending on the effects of relevant physical parameters, such as the Weber number ($We=(2\rho_{l} U^2 R) / \sigma$) and Ohnesorge number ($Oh=\mu_l / \sqrt{\rho_l R \sigma}$), where $\rho_l$ is the density of drop fluid, $\sigma$ is the coefficient of surface tension, $R$ is the radius of the drops and $U$ is the relative velocity between the drops. There are five known possible outcomes of the drop collision in the case of identical viscosity of the drops \cite{dirawi2019pof}. These are (a) slow coalescence, (b) bouncing, (c) fast coalescence, (d) reflexive separation, and (e) stretching separation. Bouncing of the drops occurs when the drops do not have enough inertia to overcome the surface tension force, while coalescence happens if the momentum of the drops is high enough to overcome the surface tension forces. In case the momentum of the drops after coalescence is high enough, then the liquid packet pulls apart into two parts, forming a dumbbell shape and the two liquid masses are connected by a cylindrical ligament at its two ends. This ligament stretches out and breaks due to the outward moment of the liquid packets, giving rise to two drops. The separation of the ligament may give rise to a small-sized satellite drop. The formation of a satellite drops does not occur as we move away from the symmetrical flow conditions in binary collision \cite{huang2019prl}. Ref. \cite{chaitanya2021pof} reported the numerical results of the collision dynamics of drops in a gaseous environment. They observed that they underwent permanent coalescence when two unequal-sized drops collided at intermediate angles ($0^\circ<\theta<23^\circ$). However, in head-on collisions and collisions with large angles ($\theta>23^\circ $), the drops exhibited reflexive separation and stretching separation, respectively. Contrary to reflexive separation, the stretching separation occurs in an oblique collision of the drops where relatively smaller volumes of the drops interact with each other \cite{chaitanya2021pof}.

Initial efforts were made to study the binary collision of drops by Qian and Law \cite{qianlaw}, who presented the regime map to distinguish between bouncing, coalescence and separation behaviours. They identified five distinct outcomes from their experiments by considering both head-on and oblique collisions between water and hydrocarbon drops of equal size. They are (i) coalescence with minor deformation, (ii) bouncing, (iii) coalescence with significant deformation, (iv) coalescence followed by separation in near head-on collisions, and (v) coalescence followed by separation in off-centre/oblique collisions. Subsequently, Ref. \cite{pan2008bouncing} reported a comprehensive investigation on the head-on collision of two identical drops, considering a wide range of Weber numbers. They employed both experimental and numerical approaches in their study. The experiments involved a time-resolved microphotographic technique similar to the method used by Qian and Law \cite{qianlaw}. Additionally, numerical simulations were conducted using a front tracking method \cite{unverdi1992front}. The researchers demonstrated that computational analysis can accurately determine the precise moment of drop merging. They achieved this by incorporating an augmented van der Waals force and an empirically derived Hamaker constant obtained from experimental observations. The study differentiated between ``soft'' collisions, resulting in minor deformations, and ``hard'' collisions, leading to significant deformations. Nobari et al. \cite{nobari1996head} also examined the head-on collision of drops of equal size. They utilized a front tracking/finite difference technique and investigated the boundaries that separate coalescing collisions from separating collisions in the Reynolds number and Weber number plane. A recent study by Deka et al. \cite{hiranya2023} for the binary collision of two non-evaporating drops shows the effect of viscosity ratio on the formation of satellite drop in reflexive separation. They have also modified the regime map in $We-Oh$ space for the collision outcomes of different viscosity drops. The above-mentioned studies exclusively examined the collision dynamics between two drops without considering evaporation. However, it is important to note that the evaporation phenomenon commonly arises in numerous industrial applications and natural phenomena \cite{ranz1952evaporation,schlottke2008direct,tripathi2015evaporating}.

Several researchers have examined the evaporation process of a single stationary or migrating drop, and their findings have been summarized in Refs. \cite{ayyaswamy,jog1996evaporation}. Subsequently, numerical simulations were carried out to investigate the dynamics of evaporating drops at various temperatures \cite{tanguy2007level,schlottke2008direct}. Tripathi and Sahu \cite{tripathi2015evaporating} utilized the evaporation model reported in Ref. \cite{schlottke2008direct} and performed three-dimensional numerical simulations using Basilisk, a volume-of-fluid-based flow solver, to study the behaviour of falling drops undergoing evaporation. By incorporating an immersed boundary method, Lupo et al. \cite{lupo2019immersed} investigated the evaporation of drops in laminar and turbulent gas flows.  

To the best of our knowledge, the interaction of drops undergoing evaporation remains largely unexplored despite its relevance in numerous practical applications. In this current study, we employ a Coupled Level Set and Volume of Fluid (CLSVOF) method \cite{sussman2000coupled,gerlach2006comparison} to investigate the various outcomes of drop-drop interactions in hot environments. The coalescence can be either permanent or followed by separation, depending on the balance between inertia and surface tension forces. The mass transfer between the drops and the surrounding medium significantly influences the outcome of the collision. To ensure the reliability of our numerical approach, we extensively validate it using various benchmark problems related to single drop evaporation \citep{pal2023evap}. In this study, we investigate the evaporation process of two drops undergoing a collision, considering different impact regimes such as bouncing, coalescence, and reflexive separation.



\section{Formulation}\label{sec:formulation}
\begin{figure}[h]
\centering
\includegraphics[width=0.7\textwidth]{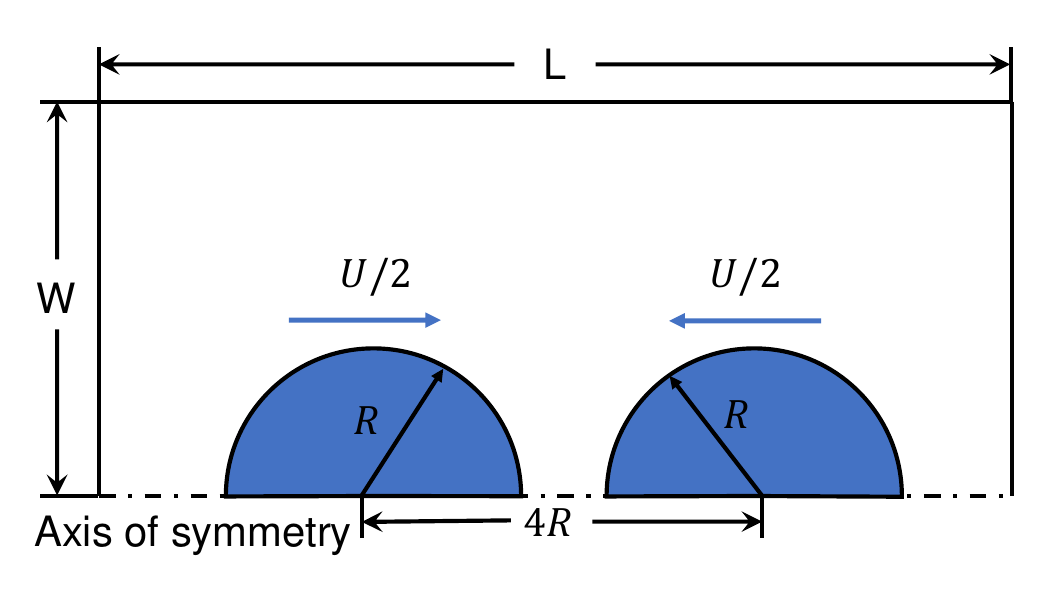}
\caption{Schematic diagram of the computational domain showing the head-on collision of two drops of radii $R_1$ and $R_2$ approaching with velocities $U_1$ and $U_2$, respectively. Here, the length $(L)$ and width $(W)$ of the computational domain are $12R_1$ and $4R_1$, respectively.}
\label{fig:fig1}
\end{figure}

The dynamics of head-on collisions of two drops in a hot ambience are investigated. Axisymmetric numerical simulations are performed using a Coupled Level Set and Volume of Fluid (CLSVOF) method in a computational domain schematically shown in Fig. \ref{fig:fig1}. We observe different collision outcomes, such as bouncing, coalescence and reflexive separation phenomena. The collisions of two drops of radii $R_1$ and $R_2$ approaching with velocities $U_1$ and $U_2$ are studied. The initial separation distance between the drops is $4R_1$, where $R_1$ is the radius of the smaller drop. The radii and velocities of the drops are considered to be equal for the collision of identical drops.

\subsection{Governing equations}
The dynamics of drop collision is governed by the solution of the Navier-Stokes equations and the interface capturing technique to resolve the evolution of the interface. To observe the evaporation of the colliding drops, the solutions of thermal energy and vapor mass transfer equations are required. For evaporation problems, the divergence-free continuity equation does not remain valid at the interface due to the phase change of the fluids. Following the one fluid formalism, the modified form of the continuity and momentum equations incorporating the source term for phase change can be written as \cite{brandt2020jcp}
\begin{eqnarray}
 \nabla \cdot \textbf{u} = \dot{m}\left(\frac{1}{\rho_g} - \frac{1}{\rho_l} \right)\delta_s,~~~~~~~~~~~~~~~~~~~~~~~~~ \label{eqn:eq1}\\
\rho (\phi) \left[\frac{\partial \textbf{u}}{\partial t} + \textbf{u}. \nabla \textbf{u} \right] = -\nabla p + \nabla \cdot [\mu(\phi)(\nabla \textbf{u} + \nabla \textbf{u}^T)] + \sigma \kappa \hat{n} \delta_s.
\label{eqn:eq2} 
\end{eqnarray}
Here, $\rho_l$ and $\rho_g$ are the densities of the liquid and gas fluids, respectively. $\textbf{u}$ is the velocity field with $u$ and $v$ as the radial and axial velocity components, respectively. $\dot{m}$ is the mass flux of the vapor generated due to evaporation. $p$ denotes the pressure field, $\sigma$ represents surface tension, $\kappa$ is the mean curvature of the interface, $\hat{n} $ is the unit normal vector on the interface, and $\delta_{s} = |\nabla F|$ is the interface delta function which is zero elsewhere except in the two-phase cells. The surface tension force is included as a volumetric force term, modelled by the continuum surface force model \cite{brackbill1992continuum}. 

The thermal energy transport and vapor transport equations are written in the following form \cite{brandt2020jcp}
\begin{eqnarray}
\rho c_p \left[\frac{\partial T}{\partial t} + \textbf{u} \cdot \nabla T \right] = k{\nabla}^2T-\dot{m}\left[h_{lg}+(c_{pl}-c_{pg})(T_{sat}-T_{\Gamma}) \right]\delta_s, \label{eqn:eq3} \\
\frac{\partial Y}{\partial t} + \textbf{u} \cdot \nabla Y = D_{lg}{\nabla}^2Y.~~~~~~~~~~~~~~~~~~~~~~\label{eqn:eq4} 
\end{eqnarray}
Here, $c_p$ is the specific heat of the fluid, $T$ and $Y$ are the temperature and vapor mass fraction field in the domain, respectively. $h_{lg}$ and $D_{lg}$ represent the latent heat of vaporization of the liquid and the binary diffusion coefficient of the vapor in the gas, respectively. $T_{sat}$ and $T_{\Gamma}$ denote the saturation and interface temperatures, respectively. The last term in the energy equation accounts for the energy that is spent in the phase change of a liquid into vapor.

The volume of fluid method is coupled with the level set method for a smooth representation of the interface. The level set function $\phi$ is the normal distance from the interface defined as
\begin{equation}
 \phi =
    \begin{cases}
     -d, & \text{in the liquid region,}\\
     0, & \text{at the interface,}\\
     +d, & \text{in the gaseous region}
    \end{cases}       
\end{equation}

In the CLSVOF interface capturing technique, the dynamics of the interface is captured by advecting the governing equations for volume of fluid and level set function, which are given by
\begin{eqnarray}
    \frac{\partial F}{\partial t} + \textbf{u} \cdot \nabla F &=& 0, ~ {\rm and}
    \label{eqn:eq6} \\ 
    \frac{\partial \phi}{\partial t} + \textbf{u} \cdot \nabla \phi &=& 0. \label{eqn:eq7}
\end{eqnarray}
For efficient calculation of mass transfer, we use the evaporation model based on an interface velocity construction step which involves solving a Poisson equation of Stefan flows. The velocity of the liquid phase is extended in the gaseous to calculate the velocity of the interface $\textbf{u}^{\Gamma}$ (see Ref. \cite{brandt2020jcp} for a detailed procedure). The interface is advected using the constructed interface velocity
\begin{eqnarray}
    \frac{\partial F}{\partial t} + \textbf{u}^{\Gamma} \cdot \nabla F &=& 0, \label{eqn:eq8} \\ 
    \frac{\partial \phi}{\partial t} + \textbf{u}^{\Gamma} \cdot \nabla \phi &=& 0. \label{eqn:eq9}
\end{eqnarray}
The normal and the curvature are calculated from the level set
function $\phi$ as
\begin{eqnarray}
    \bm{n} &=& \frac{\nabla \phi}{|\nabla \phi|},\\
    \kappa &=& -\nabla \cdot \bm{n}  = -\nabla \cdot \frac{\nabla \phi}{|\nabla \phi|}.
\end{eqnarray}
The properties near the interface are calculated using the smoothed Heaviside function. The heaviside function based on the level set function is given by, 
\begin{equation}
  H(\phi) =
    \begin{cases}
     1, & {\rm if}  \epsilon > \epsilon,\\
     \frac{1}{2}+\frac{\phi}{2\epsilon}+\frac{1}{2\pi}\left[sin(\frac{\pi\phi}{\epsilon})\right], & {\rm if} \ \phi \leq \epsilon,\\
     0, & {\rm if} \ \phi < -\epsilon.
    \end{cases}       
\end{equation}
Here, $\epsilon$ is the numerical thickness of the interface. The density, $\rho(\phi)$, the dynamic viscosity, $\mu(\phi)$, and the thermal conductivity, $k(\phi)$ are calculated in each cell from the smoothed Heaviside function using the arithmetic mean and harmonic averaging is used to calculate the specific heat \citep{brandt2020jcp},
\begin{eqnarray}
    \rho(\phi) &=& \rho_lH(\phi) + \rho_g(1-H(\phi)),  \\
    \mu(\phi) &=& \mu_lH(\phi) + \mu_g(1-H(\phi)),    \\ 
     k(\phi) &=& k_lH(\phi) + k_g(1-H(\phi)), \\
     \frac{1}{\rho c_p} &=& \frac{H(\phi)}{\rho_l c_{pl}} + \frac{1 - H(\phi)}{\rho_g c_{pg}}.
\end{eqnarray}
The mass flux at the interface $\dot m$ is calculated using the jump in vapor mass fraction at the interface \cite{irfan2017evap}, which takes the following form for a single-component liquid,
\begin{equation}
\label{eqn:eq17}
\dot m = \frac{\rho_g D_{lg}}{1-Y_{\Gamma}} \nabla_{\Gamma} Y \cdot \hat{n}  
\end{equation}
Here, $\nabla_{\Gamma} Y$ is the gradient of the vapor mass fraction at the interface and $Y_{\Gamma}$ is approximated at the interface using the following expression,
\begin{equation}
\label{eqn:eq18}
Y_{\Gamma} = \frac{p_{\Gamma}^{sat} M_l}{(p_t - p_{\Gamma}^{sat})M_g + p_{\Gamma}^{sat} M_l}.
\end{equation}
In the above equation, $p_{\Gamma}^{sat}$ and $p_t$ are the saturation pressure at interface temperature and total cell pressure, respectively. $M_l$ and $M_g$ are the molar mass of the vapor and molar mass of the gas, respectively. We use Clausius-Clapeyron equation to calculate $p_{\Gamma}^{sat}$,
\begin{equation}
\label{eqn:eq19}
p_{\Gamma}^{sat} = p_t \exp\left[-\frac{h_{lg}M_l}{R}\left(\frac{1}{T_{\Gamma}}-\frac{1}{T_{sat}}\right)\right].
\end{equation}

The various dimensionless numbers associated with the problem considered in the present study are the Weber number ($We=2\rho_lU^2R_1/\sigma$), Ohnesorge number ($Oh=\mu_l/\sqrt{\rho_lR_1\sigma}$), Stefan number ($St=c_{pg}(T_\infty-T_d)/h_{lg}$), and radius ratio $(R_r=R_2/R_1$). Here, the Stefan number $(St)$ represents the measure of the level of superheat between the drop and surrounding, and $c_{pg}$ is the isobaric specific heat of the gas phase, $h_{lg}$ is the latent heat of vaporization of the liquid, $T_d$ and $T_\infty$ are the initial temperatures of the drop and the surroundings, respectively. In the present study, we have investigated the collision dynamics of two identical evaporating drops $(R_r=1)$, except in the validation section, where we study unequal-sized drops. In all the simulations, we consider $U_1=U_2$, such that, the relative velocity, $U=(U_1+U_2)/2$. The dimensionless time is given by $\tau=t/t_i$, where $t_i=2R_1/U$ represents the collision time scale.

\subsection{Numerical method} 
The governing equations are discretized using a finite difference approach. We have followed a Coupled Level Set and Volume of Fluid (CLSVOF) method \cite{gerlach2006comparison} that combines the advantages of the level-set (LS) \cite{sussman1998improved} and volume-of-fluid (VOF) \cite{welchnwilson} methods. A staggered grid (MAC) arrangement \cite{harlow1965numerical} is employed in our simulations. In such a grid arrangement, the scalar quantities such as pressure, temperature and vapor mass fraction are defined at the cell centers and the vector quantities, such as the velocity components, are described at the center of the cell faces to which they are normal. The grid size in the radial and axial directions is considered to be the same, i.e., $\Delta r = \Delta z$. The convective terms in the momentum equation are discretized using the higher order essentially non-oscillatory (ENO) scheme as described in Refs. \cite{osher88} and \cite{chang1996level} while a second order central difference scheme is used to discretize the viscous terms. The surface tension term in the momentum is handled using the continuum surface force model of Ref. \cite{brackbill1992continuum}. The discretized form of the momentum equation is advanced in time explicitly, thus, obtaining a provisional velocity field. Such a velocity field may not be divergence-free since it does not satisfy the continuity equation in each cell. The compliance of the continuity equation is achieved by solving the corresponding Poisson equation for pressure correction. The pressure correction equation is solved using the open-source algebraic multi-grid solver HYPRE \cite{hypre}. The previously obtained provisional velocity field is then corrected using the pressure correction values. Thus, the converged solution is obtained at a new time level after achieving a divergence-free velocity field.

After obtaining the flow velocity field, calculations for the interface velocity field are performed. To calculate the interface velocity, we first solved the following Poisson equation to calculate the Stefan flow \citep{brandt2020jcp},
\begin{equation}
\label{eqn:eq20}
\nabla^2 \psi = \dot{m}\left(\frac{1}{\rho_g} - \frac{1}{\rho_l}\right)\delta_s.
\end{equation}

The Stefan flow is given by $\textbf{u}^s = \nabla \psi$. The Stefan flow velocity is subtracted from the flow velocity to obtain the divergence-free extended liquid velocity ($\textbf{u}^e$), which is non-zero only in the gaseous domain. Eventually, the interface velocity $\textbf{u}^{\Gamma}$ is calculated as (for detailed see Ref. \cite{brandt2020jcp})
\begin{equation}
\label{eqn:eq21}
 \textbf{u}^{\Gamma} = \textbf{u}^e - \frac{\dot{m}}{\rho_l}n.
\end{equation}

The advection equations of the volume fraction and level-set function are advected with the constructed interface velocity $\textbf{u}^{\Gamma}$ to obtain the new volume fraction field $F^{n+1}$ and the level-set function $\phi^{n+1}$, which are essential for interface reconstruction. The second-order conservative operator split advection scheme \cite{rudman1997} is used for the discretization of the volume fraction advection equation (Eq. \ref{eqn:eq8}). In order to obtain higher accuracy, divergence correction is implemented at the interface \cite{gerlach2006comparison,puckett1997high,rider1998reconstructing}. Thus, Eq. (\ref{eqn:eq8}) is reformulated into the conservative form along with the implementation of divergence correction as ${\partial F / \partial t + \nabla \cdot (F\textbf u^{\Gamma})= F\nabla \cdot \textbf u^{\Gamma}}$ which is then solved using the operator split advection scheme. The conservation of $F$ is maintained by employing an implicit scheme in the first sweeping direction and an explicit scheme in the second sweeping direction, as suggested by Puckett et al. \cite{puckett1997high}. The approach is made second-order accurate by alternating the sweep directions in each time step, commonly known as {\it Strang splitting} \cite{strang1968construction}. The level-set advection equation (Eq. \ref{eqn:eq9}) is simultaneously solved in the corresponding directions by discretizing the convective terms using the ENO scheme. At each time step, after finding the updated volume fraction $F^{n+1}$ and level set function $\phi^{n+1}$, the level set function is reinitialized to maintain the exact signed normal distance from the reconstructed interface by coupling the level set function with volume fraction \cite{sussman2000coupled,son2002coupled,son2003efficient}.

The energy and vapor mass transport equations are solved with the flow velocity field $\textbf{u}$. The temporal term in the energy equation is discretized using a second-order accurate scheme, while first-order explicit discretization is used for the vapor transport equation. The convective terms of both the equations are discretized using the third-order accurate QUICK \cite{leonard} scheme, and second-order central differencing is used for the diffusion terms. In the present work, the time-stepping procedure is based on an explicit method to maintain the stability of the solution. Five stability conditions are associated with the set of governing equations. The outcome of such stability conditions is convective time, momentum diffusion time, capillary time, thermal diffusion time, and vapor mass diffusion time. The time step for the calculations is chosen to be the smallest of the above.

\section{Validation}\label{sec:validation}

To verify the accuracy of our numerical technique, we first simulated the scenarios previously studied by other researchers involving the collision of non-evaporating drops, which were investigated both experimentally and numerically. Subsequently, we performed a grid convergence test for the collision of two evaporating drops associated with a typical set of parameters considered in the present study.

\subsection{Collision in bouncing regime} \label{subsec:bouncing_drops}

\begin{figure}[ht]
\centering
\includegraphics[width=0.95\textwidth]{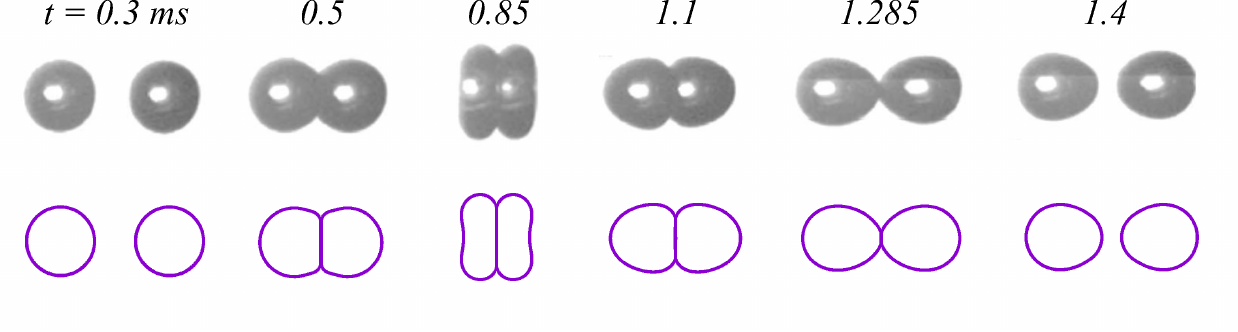}
\caption{Temporal evolution of two identical drops ($R_1=R_2=167.6$ $\mu$m) undergoing bouncing separation phenomenon. The dimensionless parameters used for the simulation are $We = 9.706$ and $Oh = 0.037$. The top and bottom rows show the experimental results of Ref. \cite{pan2008bouncing} and the results obtained from the present computation.}
\label{fig:fig2}
\end{figure}

The collision dynamics of two non-evaporating tetradecane drops in the air in the bouncing regime is simulated, and the results obtained through our simulations are compared with the experimental results \citep{pan2008bouncing} in order to validate our predictive procedure. The collision of two identical drops of radius $R_1=R_2=167.6$ $\mu$m with a relative velocity of $U=0.992$ m/s between them is simulated in an axisymmetric domain \ref{fig:fig1}. Table \ref{tab:table1} presents the physical parameters used for the simulations. Zero pressure outflow boundary conditions and zero gradient velocity boundary conditions are applied at all the confining boundaries except at the axis of symmetry, where symmetric boundary conditions are applied for both pressure and velocity. The standard interface capturing techniques, such as volume of fluid and level set methods, the interfaces coalesce numerically as soon as two interfaces are presented in a single computational cell. In our simulations, only half of the domain (as shown in Fig. \ref{fig:fig1}) is considered with the application of Ghost cell boundary condition \cite{jiang2007gcm} at the mirror axis. In this case, $We=9.629$ and $Oh=0.0356$. The comparison of the shapes of the drops at different time instants are shown in Fig. \ref{fig:fig2} and it is evident that the results obtained through present simulations are in excellent agreement with the numerical and experimental results of Pan et al. \cite{pan2008bouncing}. The drops undergo significant deformation as they approach each other and finally bounce back. It can be seen that our technique captures well the shape deformation of the drops.

\begin{table}[ht]
\caption{Properties of the fluids used in our simulations. The values of the surface tension $(\sigma)$, latent heat of vaporization $(h_{lg})$ and binary diffusion coefficient of the vapor in the gas at the air ($D_{lg}$) for the tetradecane-air interface are $26 \times 10^{-3}$ N/m, $352.8$ kJ/kg and $2.0 \times 10^{-5}$ $m^2$/s, respectively.}\label{tab:table1}
\begin{tabular}{|c|c|c|c|c|c|}
\hline
Fluids & $\rho$  & $\mu$  & $c_p$  & $k$ & $M$   \\
 & (kg/m$^3$) & (Pa$\cdot$s) & (J/kgK) &(W/mK) & (kg/kmol)  \\
\hline
Tetradecane & $759.0$ & $2.05 \times 10^{-3}$ & $2.213 \times 10^{3}$ & $131.7 \times 10^{-3}$ & $142.2$  \\ \hline
Air &$1.2$ & $1.78 \times 10^{-5}$ & $1.013 \times 10^{3}$ & $33.6 \times 10^{-3}$ & $29.0$\\ \hline
\end{tabular}
\end{table}

\subsection{Collision in reflexive separation regime}

\begin{figure}[ht]
\centering
\includegraphics[width=0.95\textwidth]{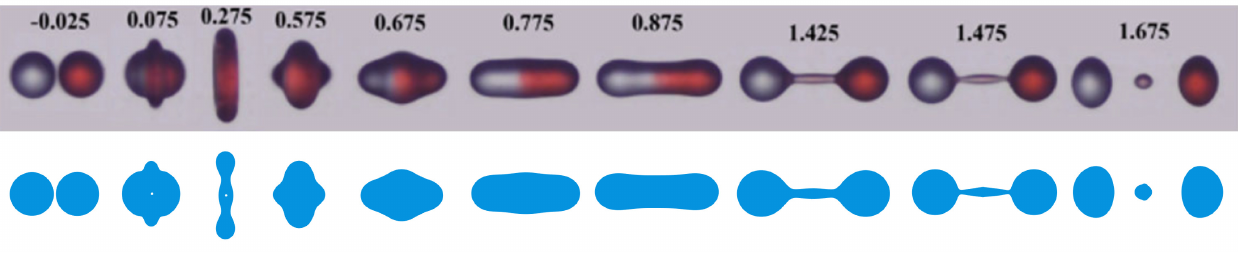}
\caption{Sequence of collision of two tetradecane drops with $R_1=R_1 = 150$ $\mu$m in air. The values of the dimensionless parameters are $We=45.92$ and $Oh=0.0376$. The top and bottom rows show the experimental results of Ref. \cite{huang2019prl} and the results obtained from the present computation.}
\label{fig:fig3}
\end{figure}

\begin{figure}[ht]
\centering
\includegraphics[width=0.9\textwidth]{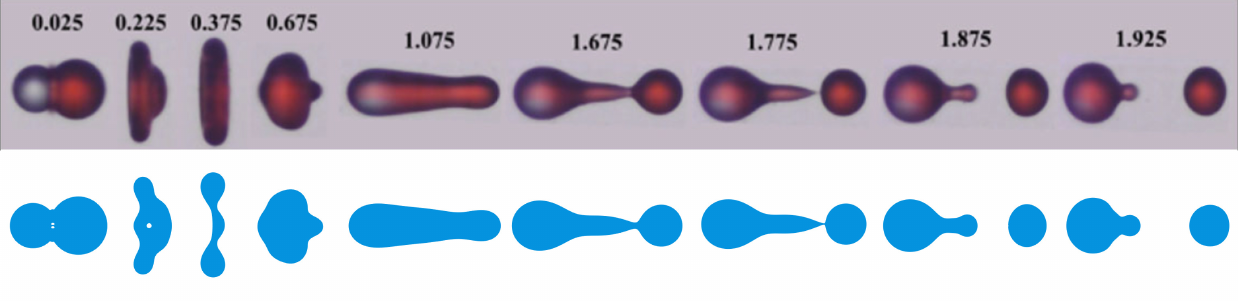}
\caption{Sequence of collision of two unequal-sized tetradecane drops in the air. Here, $R_1=150$ $\mu$m and $R_2 = 192$ $\mu$m. The values of the dimensionless parameters are $We=46.32$ and $Oh=0.0376$. The top panel shows the results obtained by Huang et al. \cite{huang2019prl}, and the bottom panel shows the results due to present computations.}
\label{fig:fig4}
\end{figure}

The collision of two non-evaporating tetradecane drops ($R_1=R_2=150$ $\mu$m) in the air is simulated in the reflexive separation regime. The computational domain remains the same as shown in Fig. \ref{fig:fig1} with $W=4R_1$ and $L=12R_1$. The drops are initialized at a distance of $4R_1$ from each other. The zero pressure outflow boundary condition and zero gradient velocity boundary condition are applied at all the confining boundaries, and the symmetric boundary conditions are applied at the axis of symmetry. The properties of the fluids are taken the same as for the earlier case (Table \ref{tab:table1}). Figure \ref{fig:fig3} shows the evolution of identical drops as they collide with each other and separate. It can be observed that the liquid packet formed after collision elongates due to the effect of high inertia, and a thin ligament is formed before the pinch-off takes place due to the effect of capillary forces. Due to the symmetry in the initial shape of the drops, the pinch-off forms the two extreme ends of the ligament, giving rise to the formation of a small-size satellite drop, as explained by Huang et al. \cite{huang2019prl}. The outcomes of the collision of the drops with different initial radii ($R_1=150$ $\mu$m, $R_r=1.28$) are shown in Fig. \ref{fig:fig4}. For this case, the collision is no longer symmetric; thus, the satellite drop formation does not occur \cite{hiranya2023}.

\subsection{Grid independence study}
\begin{figure}
\centering
\includegraphics[width=0.95\textwidth]{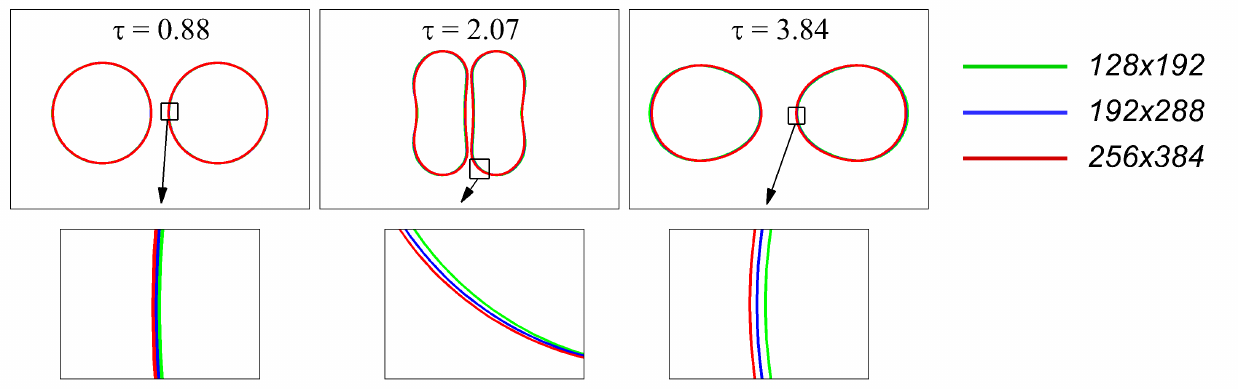}
\caption{Grid convergence test. This figure compares the interface of the drops obtained using different grids at three different time instances. The bottom row shows the magnified views of the interface in the region marked by the square box in the top row. The dimensionless parameters used for the simulation are $We=9.706$, $Oh=0.037$, and $St=0.302$.}
\label{fig:fig5}
\end{figure}

\begin{figure}
\centering
\includegraphics[width=0.7\textwidth]{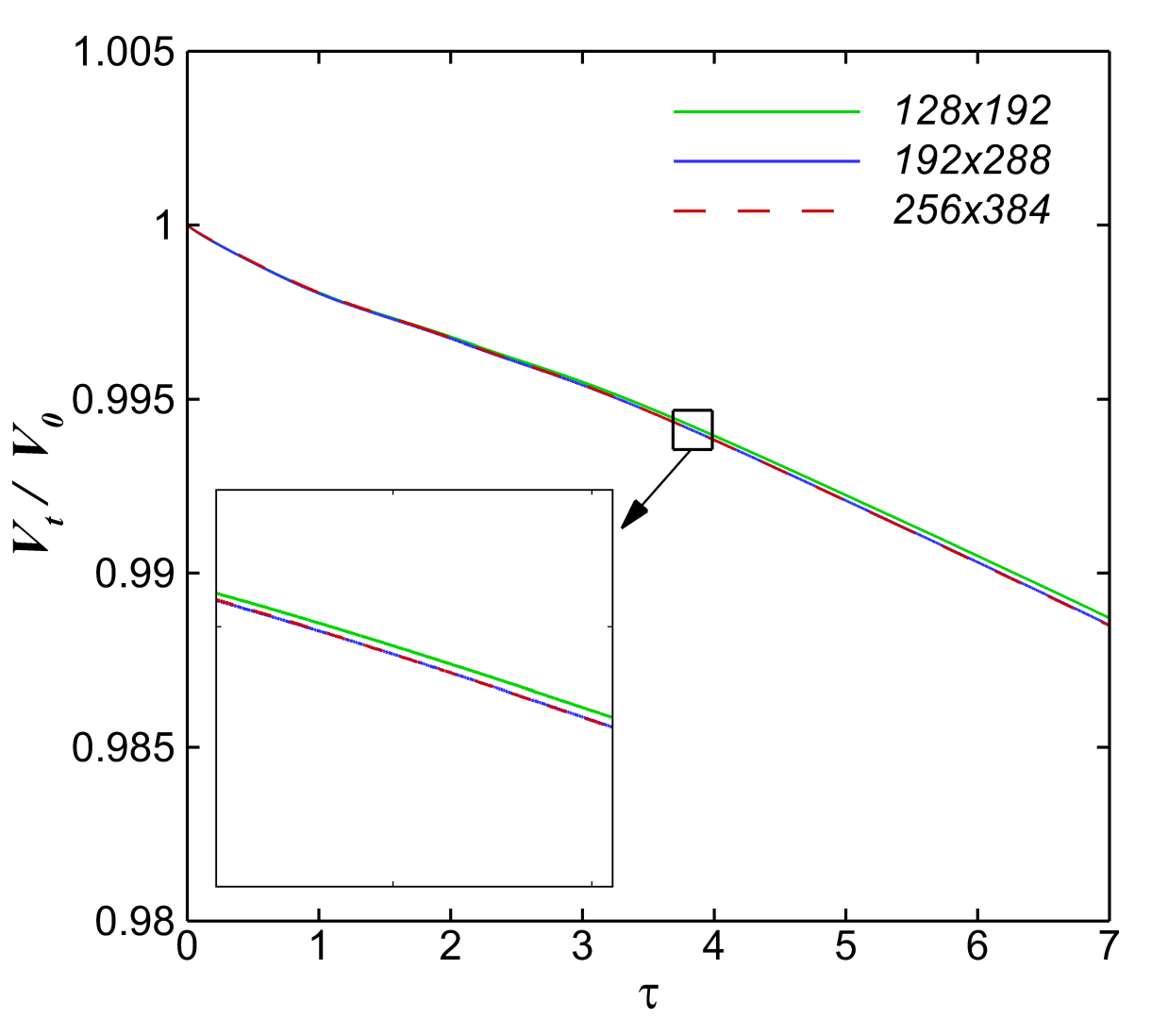}
\caption{Temporal evolution of dimensionless liquid volume for different grids considered. The dimensionless parameters used for the simulation are $We=9.706, Oh=0.037$, and $St=0.302$.}
\label{fig:fig6}
\end{figure}

A grid independence test was conducted to determine the adequate grid size for the numerical calculations present in this document. Figure \ref{fig:fig5} shows the case of two identical evaporating drops interacting in the bouncing regime for the Stefan number, $St=0.302$. The values of the rest of the dimensionless numbers are $We=9.706$ and $Oh=0.037$. The computational configuration and the boundary conditions are kept the same as for the case considered in section "Collision in bouncing regime". The Dirichlet boundary condition for temperature is applied at all domain boundaries except the axis of symmetry to maintain the level of superheat. For the grid independence test, three different uniform mesh configurations were considered, namely $128 \times 192$ (with a uniform mesh size, $\Delta=R_1/32$), $192 \times 288$ ($\Delta=R_1/48$), and $256 \times 384$ ($\Delta=R_1/64$). Here, $R_1=R_2=150$  $\mu$m. A comparison of the interface profiles of the drops obtained using different grids considered is shown in Fig. \ref{fig:fig5}. The depicted time instances correspond to when the drops approach each other ($\tau=0.88$), reach maximum deformation ($\tau=2.07$), and separate from each other ($\tau=3.84$). In order to see the effect of grid size on the mass transfer rate from the drops, we compared the time evolution of dimensionless liquid volumes in the domain (Fig. \ref{fig:fig6}). The insets in Figs. \ref{fig:fig5} and \ref{fig:fig6} show the magnified views of the plots. It can be concluded from the figures that the grid size ($\Delta = 2.343$ $\mu$m) corresponding to grid mesh $256 \times 384$ is the most appropriate, and the same grid size is adapted for all the computations presented in this manuscript. We have chosen tetradecane as the working fluid. 

\section{Results and Discussion}\label{sec:results}

We begin the presentation by quantifying the effect of evaporation on the collision dynamics of the drops. Modified forms of continuity and momentum equations are solved along with energy and vapor mass transport equations to bring into the effect of mass transfer. The drop is initialized at a temperature of $T_d$, and the ambient temperature ($T_\infty$) is initialized at a temperature higher than the drop temperature ($T_{\infty}>T_d$). The steady-state solution of the vapor transport equation without convective terms is set as the initial condition for vapor mass fraction in the domain. Dirichlet boundary conditions for temperature  ($T=T_{\infty}$) and vapor mass fraction ($Y_{\infty}=0$) are set at all the boundaries of the computational domain except the axis of symmetry where zero gradient boundary conditions are applied. Unless mentioned otherwise, the initial and boundary conditions in the domain are identical for all the cases analyzed in this section. The numerical technique has been tested rigorously for the evaporation of a single drop by considering different benchmark cases (for more details, see Ref. \cite{pal2023evap}).

\subsection{Bouncing separation with evaporation}\label{subsec:bouncing_evap}

\begin{figure}[ht]
\centering
\includegraphics[width=0.9\textwidth]{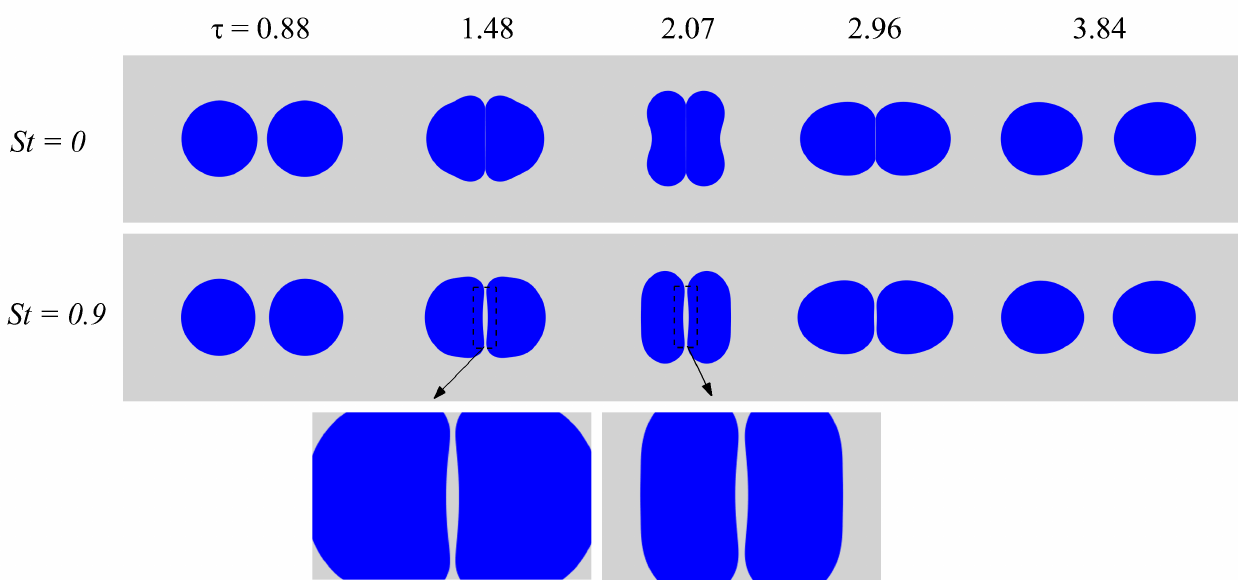}
\caption{Comparison of the collision dynamics of the drops in the absence of evaporation ($St=0$) and with the evaporation for $St=0.9$. The rest of the parameters are $R_1=R_2=167.6$ $\mu$m, $We=9.706$, and $Oh=0.037$. The magnified views depicting the region separating the drops for the evaporation case at $\tau=1.48$ and 2.07 are also shown at the bottom. Note that the computational domain considered in our simulations is large and figures display only a subset of the computational domain.}
\label{fig:fig7}
\end{figure}

\begin{figure}
\centering
\includegraphics[width=0.95\textwidth]{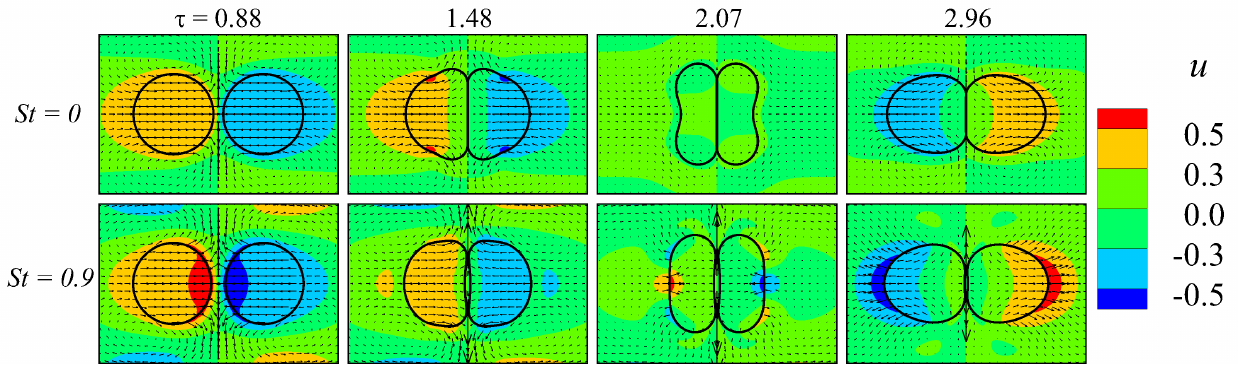}
\caption{Comparison of the velocity field during the collision of drops in the absence of evaporation ($St=0$) and with the evaporation for $St=0.9$. The rest of the parameters are $R_1=R_2=167.6$ $\mu$m, $We=9.706$, and $Oh=0.037$. Note that the computational domain considered in our simulations is large, and figures display only a subset of the computational domain.}
\label{fig:fig8}
\end{figure}

In this section, the collision dynamics of two identical drops ($R_1=R_2=167.6$ $\mu$m) colliding in the bouncing regime in the presence of a hot ambience is analyzed. Along with the flow equations (Eqs. \ref{eqn:eq1} and \ref{eqn:eq2}) and interface capturing equations (Eqs. \ref{eqn:eq8} and \ref{eqn:eq9}), the energy (Eq. \ref{eqn:eq3}) and vapor transport (Eq. \ref{eqn:eq4}) equations are also solved in axisymmetric configuration (Fig. \ref{fig:fig1}) to capture the flow physics. As mentioned earlier, the ghost cell method \cite{jiang2007gcm} has been applied, and only half of the computational domain, as shown in Fig. \ref{fig:fig1}, is simulated. In this case, the relative velocity of the drops is $U=0.992$ m/s. The values of the rest of the dimensionless numbers are $We=9.706$ and $Oh=0.037$. The effect of evaporation is brought about by varying the Stefan number ($St$), which is controlled by changing the temperature of the surrounding air. The initial temperature of the drop is kept fixed at $T_d = 0.9 \times T_{sat}$ for all the results presented in this section. The top and bottom rows of figure \ref{fig:fig7} depict the time evolution of the process of bouncing of the two drops in the absence $(St=0)$ and the presence of evaporation with $St=0.9$, respectively. In the absence of evaporation, it can be seen that the drops start deforming as they come close to each other. The interface assumes a flat shape on the side in contact with the other drop, and the opposite side keeps on deforming. The deformation of the drops is maximum at the point when the kinetic energy of the drops is transformed completely into the surface energy. \ks{The drops do not have enough energy to overcome the capillary forces. For this set of parameters, the interface rupture does not occur, and the drops remain separated by an air gap approximately equal to one grid size in thickness.} After this point, the surface tension effect tries to bring back the drops into an equilibrium shape by decreasing their surface energy. Thus, the conversion of the surface energy to the kinetic energy starts, and the liquid masses begin to pull apart, giving rise to the bouncing separation.

\begin{figure}[ht]
\centering
\includegraphics[width=0.9\textwidth]{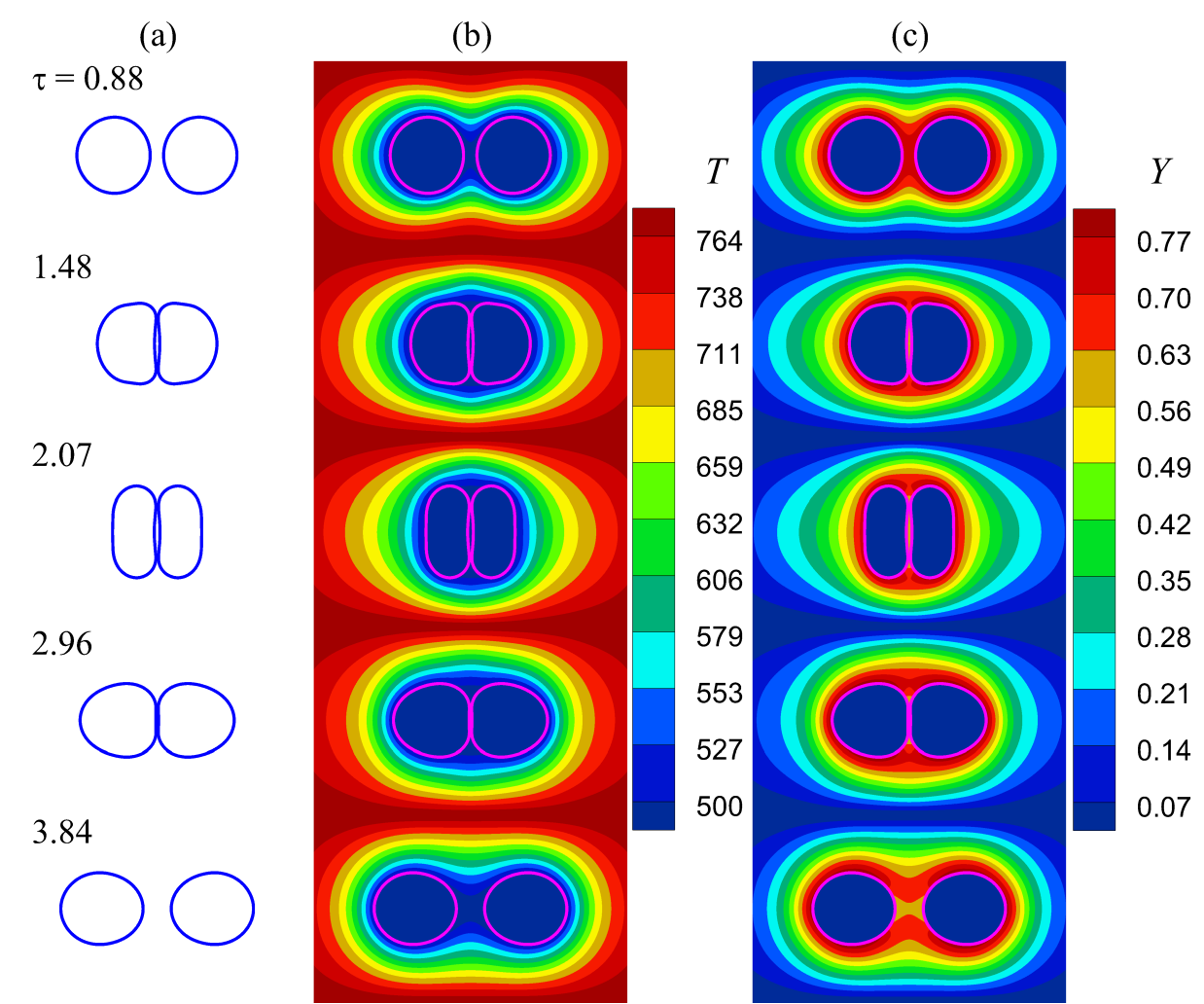}
\caption{Contours of the (a) interfaces of the drops characterised by the zero level-set function, (b) temperature field ($T$), and (c) vapor mass fraction ($Y$) for $St=0.9$. The rest of the parameters used in the simulations are $R_1=R_2=167.6$ $\mu$m, $We=9.706$, and $Oh=0.037$.}
\label{fig:fig9}
\end{figure}

\begin{figure}[ht]
\centering
\includegraphics[width=0.7\textwidth]{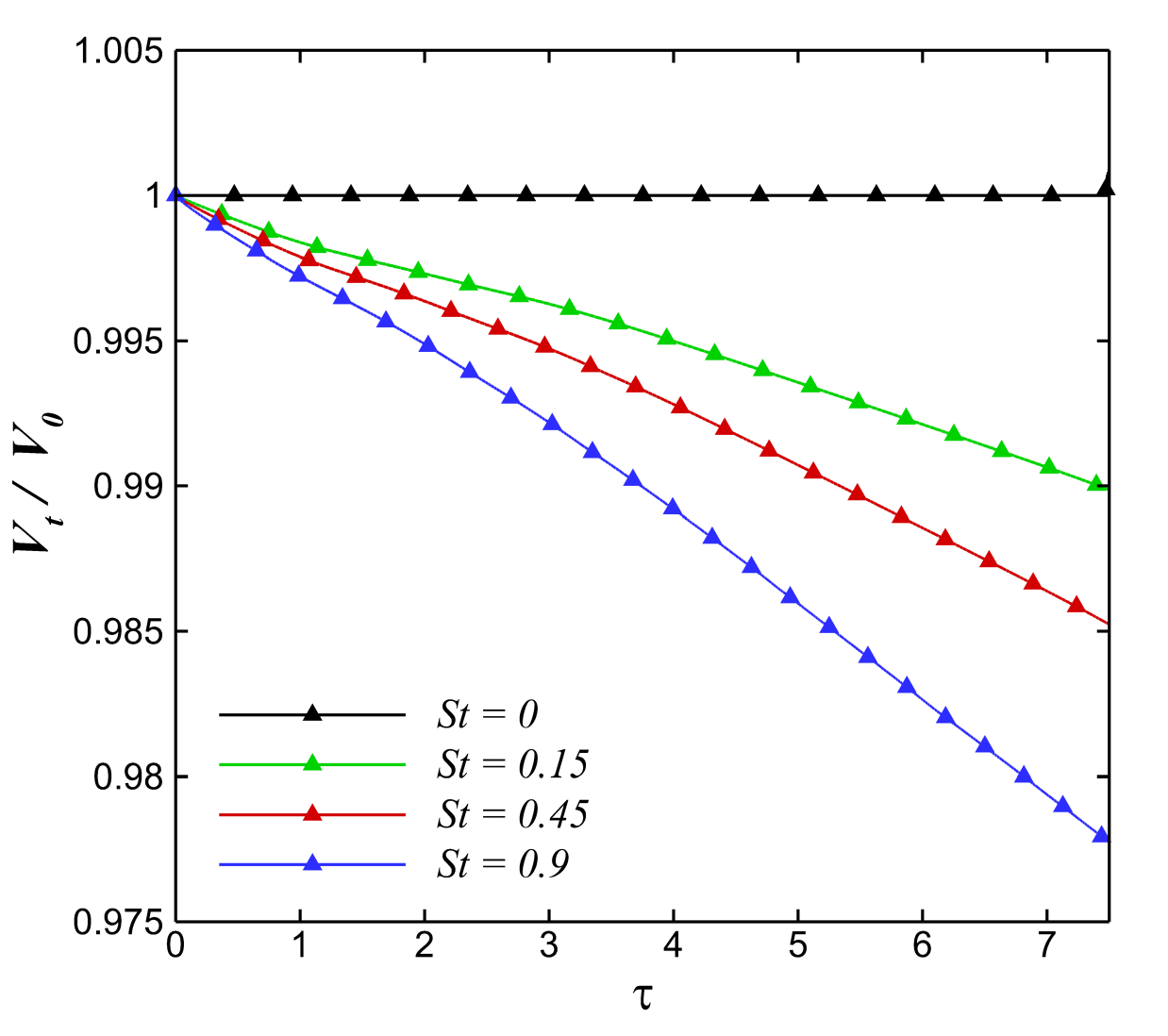}
\caption{Temporal evolution of the total liquid volume with the initial liquid volume of the drops for different values of Stefan numbers. The rest of the parameters are $R_1=R_2=167.6$ $\mu$m, $We=9.706$, and $Oh=0.037$.}
\label{fig:fig10}
\end{figure}

The effect of evaporation on the collision dynamics of the bouncing separation can be seen in the bottom row of Fig. \ref{fig:fig7}. Visual inspection of the drop shapes reveals that in the presence of evaporation, the interfaces at the approaching sides of the drops do not become flat in contrast to the case without evaporation. The interface remains deformed at the approaching sides and forms a cavity-shaped enclosure that entraps some vapor between the two drops. This is further visible in the magnified views as shown in the bottom of Fig. \ref{fig:fig7}. The velocity contours superimposed with the interface profiles for the cases with and without evaporation are demonstrated in Fig. \ref{fig:fig8}. It can be seen that in the case of evaporation, the Stefan flow generated due to phase change is radially outwards from the interfaces of the drops. This flow is responsible for the small gap between the two drops and does not let the interfaces touch each other, entrapping some vapor. The magnitude of the velocity vectors is very high in the gap since the vapor is squeezed and flows out to escape the gap (see panels 3 and 4 in the bottom row of Fig. \ref{fig:fig8}). The contours of the interface of the drops characterised by the zero level set, temperature field and the vapor mass fraction field are plotted in Fig. \ref{fig:fig9} for the maximum level of superheat considered ($St=0.9$).

\begin{figure}[ht]
\centering
\includegraphics[width=0.7\textwidth]{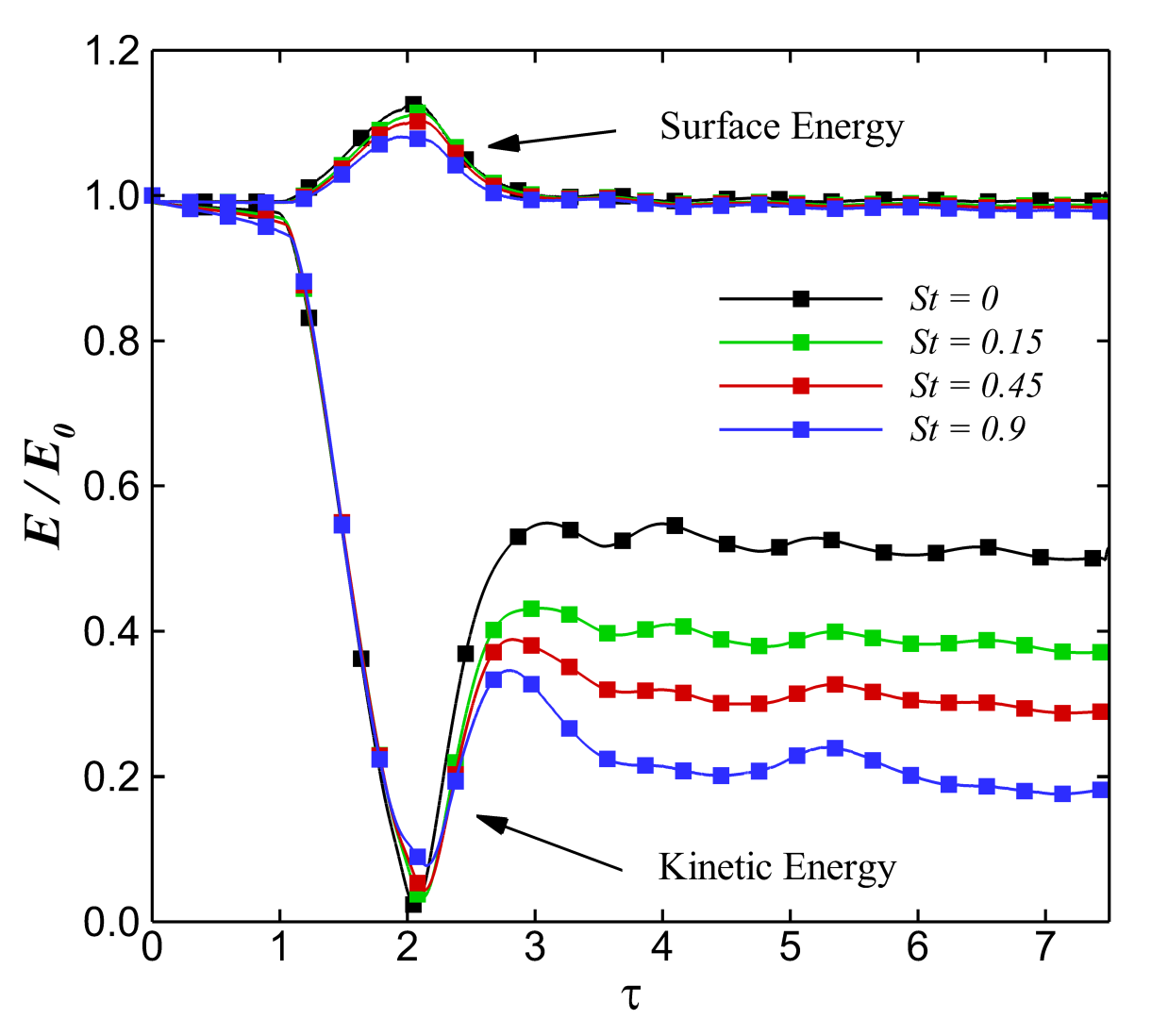}
\caption{Temporal evolution of normalized kinetic and surface energy of the drops in bouncing drops paradigm for different values of Stefan numbers. The rest of the parameters are $R_1=R_2=167.6$ $\mu$m, $We=9.706$, and $Oh=0.037$.}
\label{fig:fig11}
\end{figure}

\begin{figure}[ht]
\centering
\includegraphics[width=0.7\textwidth]{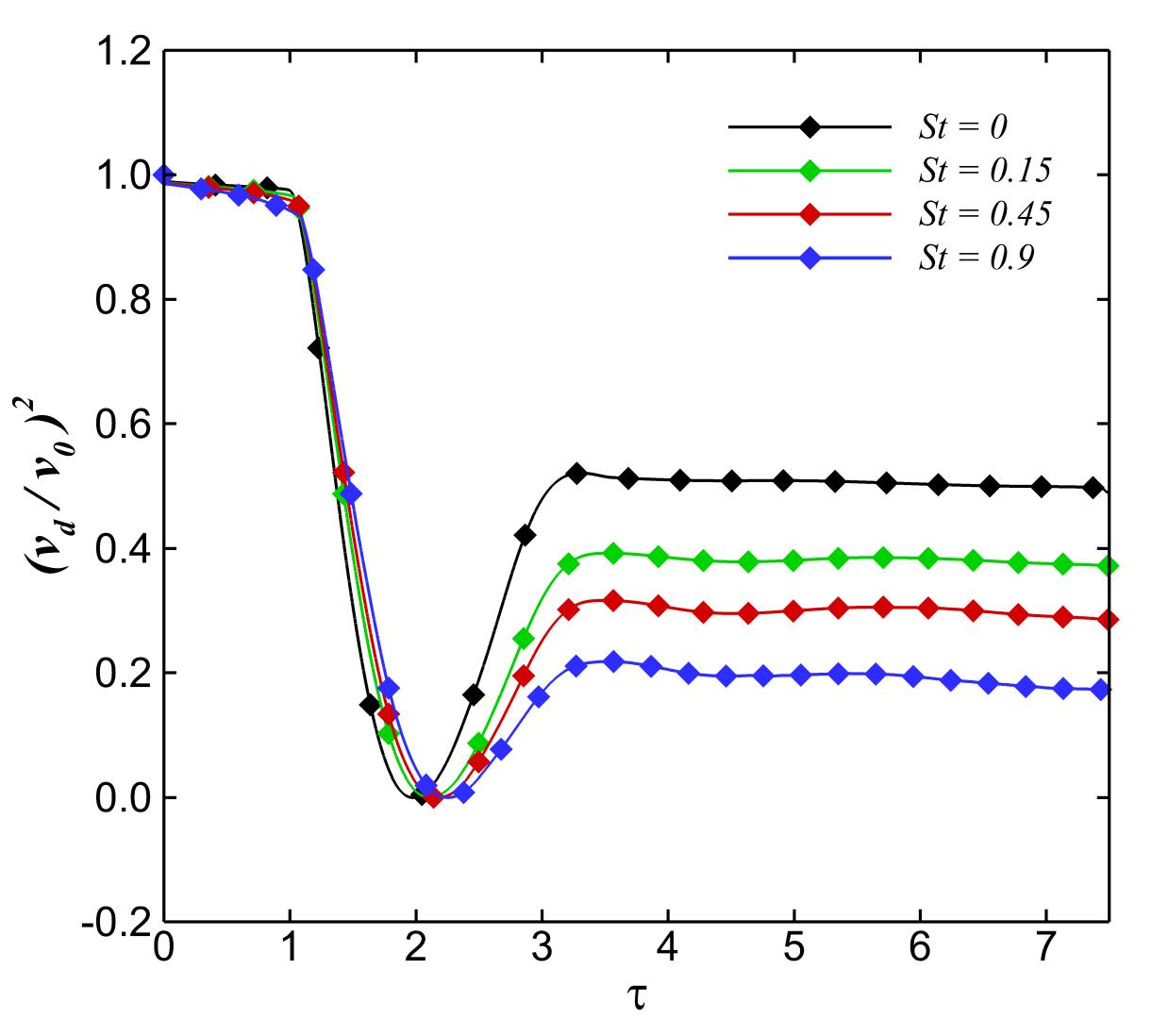}
\caption{Temporal evolution of normalized squared velocities of the drops in bouncing drops paradigm for different values of Stefan numbers. The rest of the parameters are $R_1=R_2=167.6$ $\mu$m, $We=9.706$, and $Oh=0.037$.}
\label{fig:fig12}
\end{figure}

The total volume of the liquid, non-dimensionalized with the initial liquid volume of the drops, is plotted with time for different combinations of drop and ambient temperatures (Fig. \ref{fig:fig10}) characterised by the Stefan number $(St)$. The evaporation rates of the liquid show non-linear behavior, and the mass transfer rate decrease when two drops are close to each other due to the presence of more entrapped vapor between the drops. \ks{At each time step, the total kinetic energy of the liquid is computed by summing the kinetic energy across all computational cells containing the liquid. Within each computational cell, the kinetic energy is determined using the expression: $\Delta KE = 0.5 \times F \rho_l (u^2 + v^2)dv$, where $F$ is the volume fraction of liquid, $\rho_l$ is the density of liquid, $dv$ is the volume of the computational cell, and $u$ and $v$ represent the axial and radial components of the velocity field. The surface energy is calculated as $SE = \sigma \Sigma A_s$, where $A_s$ is the total surface area of the interface. $A_s$ is computed by integrating the area of the interface across all two-phase cells.} The kinetic energy (KE) and surface energy (SE) of the drops normalized with their initial values for different levels of superheat are compared in Fig. \ref{fig:fig11}. The peak of the surface energy of the drops at the maximum deformation (corresponding to $\tau=2$) decreases with increasing evaporation. \ks{This occurs because, with increased evaporation, the total surface area of the liquid diminishes, resulting in a decrease in surface energy.} The kinetic energy of the drops decreases more after the collision, with an increase in evaporation leading to slower separation velocities of the drops. The reason for this is in two folds: one, that the liquid mass is depleted with time as a consequence of evaporation, and the second is that the velocity of the drops is decreased due to the effect of radially outward evaporation velocity. To confirm this, we have plotted the dimensionless squared velocity of the liquid in Fig. \ref{fig:fig12}. It can be observed that the separation velocity of the drops also decreases with increasing evaporation.

\begin{figure}[ht]
\centering
\includegraphics[width=0.7\textwidth]{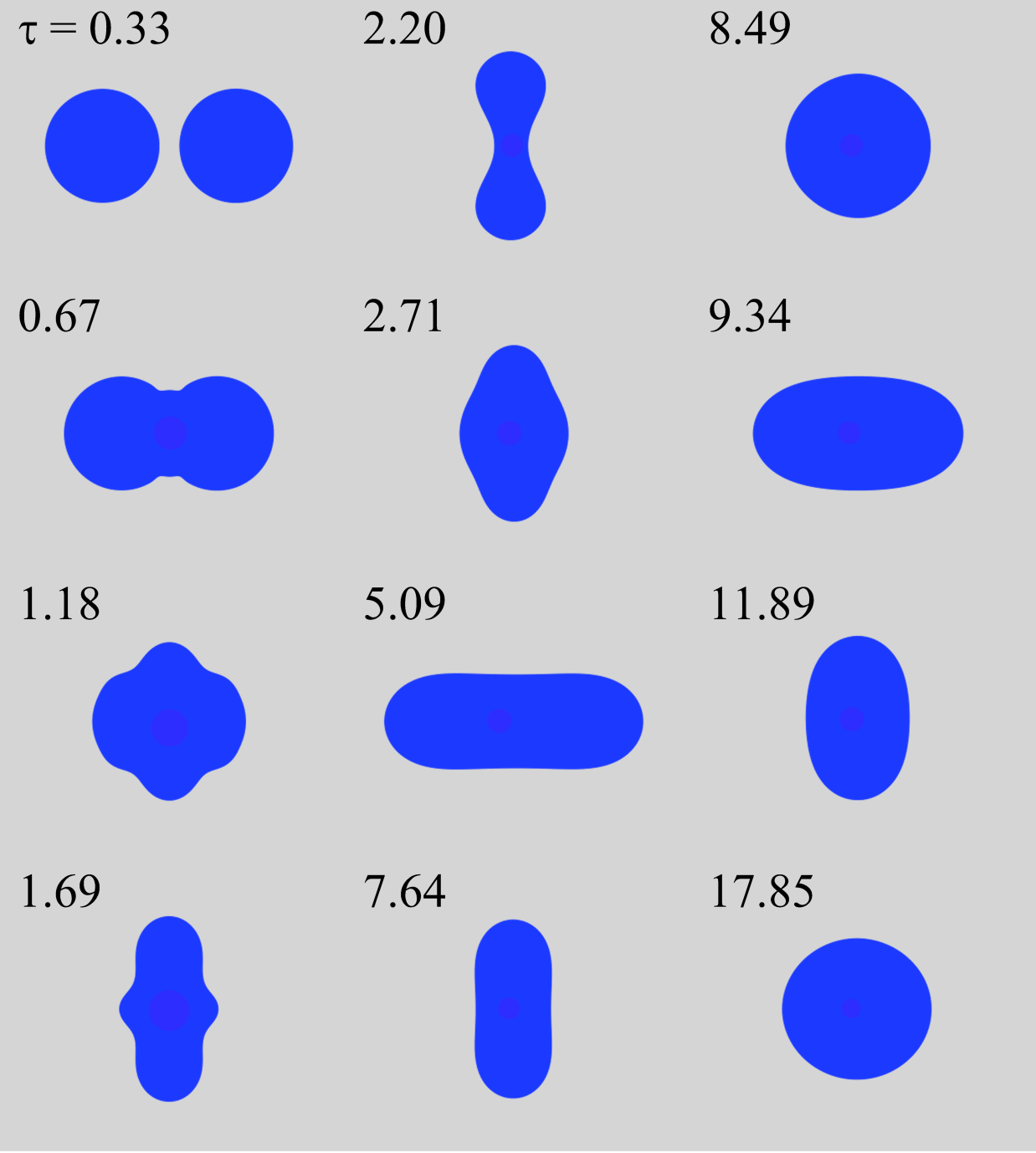}
\caption{Sequence of coalescence of two identical tetradecane drops for $R_1=R_2=169.7$ $\mu$m, $We=13.63$, $Oh=0.0354$, and $St=0$.}
\label{fig:fig13}
\end{figure}

\subsection{Coalescence with evaporation}\label{subsec:coalescence_evap}
After analyzing the effects of mass transfer in the bouncing separation of two drops, we focus on the situation related to coalescence. The coalescence of two drops occurs when the drops have sufficiently large momentum to overcome the effect of capillary forces. During the coalescence process, the interfaces of the two drops break at the line of merger, and the liquid from each drop flows into the other, forming a single conglomerate. The liquid conglomerate oscillates by elongating in the axial and radial directions alternatively. Thus by losing the kinetic energy in each oscillation due to the viscous effect, eventually, the conglomerate takes a spherical shape. During this process, continuous conversion of the surface energy to the kinetic energy and vice versa takes place.

\begin{figure}
\centering
\includegraphics[width=0.9\textwidth]{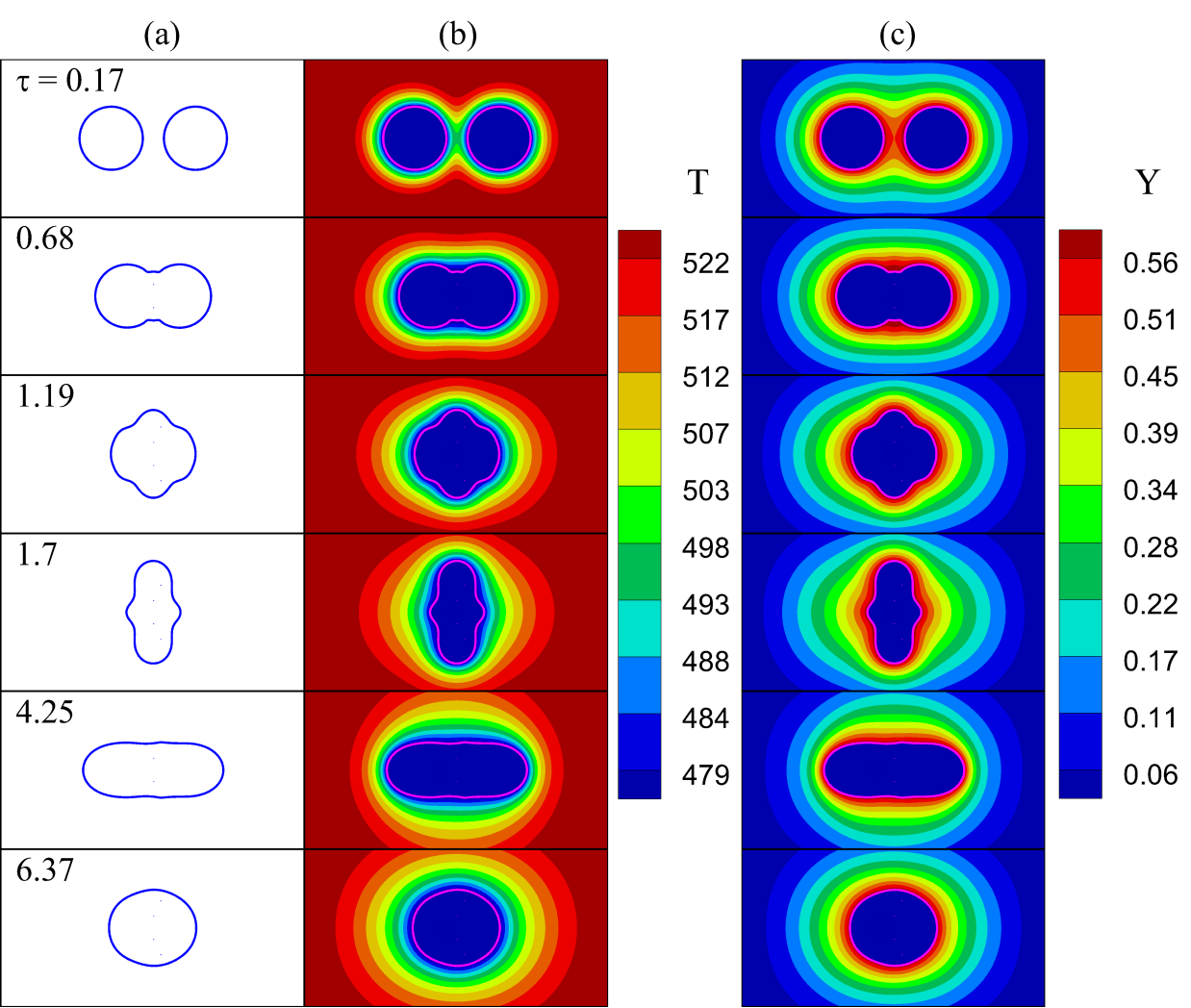}
\caption{Collision dynamics of two evaporating drops undergoing coalescence. Contours of the (a) interfaces of the drops characterised by the zero level-set function, (b) temperature field ($T$), and (c) vapor mass fraction ($Y$). The rest of the parameters are $R_1=R_2=169.7$ $\mu$m, $We=13.63$, $Oh=0.0354$, and $St=0.15$.}
\label{fig:fig14}
\end{figure}

\begin{figure}
\centering
\includegraphics[width=0.7\textwidth]{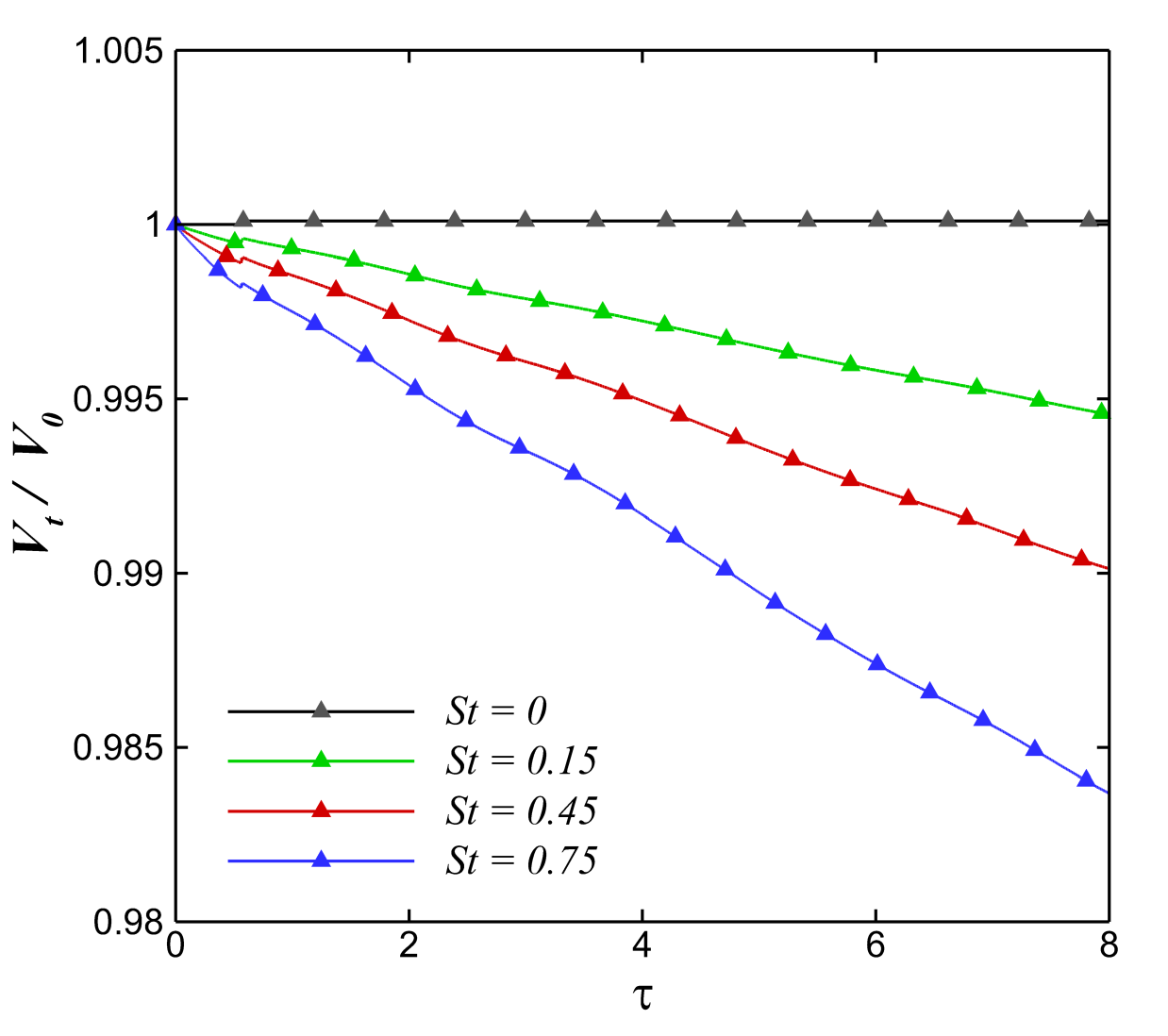}
\caption{Temporal evolution of the total liquid volume with the initial liquid volume of the drops for different values of Stefan numbers. The rest of the parameters are $R_1=R_2=169.7$ $\mu$m, $We=13.63$, and $Oh=0.0354$.}

\label{fig:fig15}
\end{figure}
\begin{figure}[ht]
\centering
\includegraphics[width=0.7\textwidth]{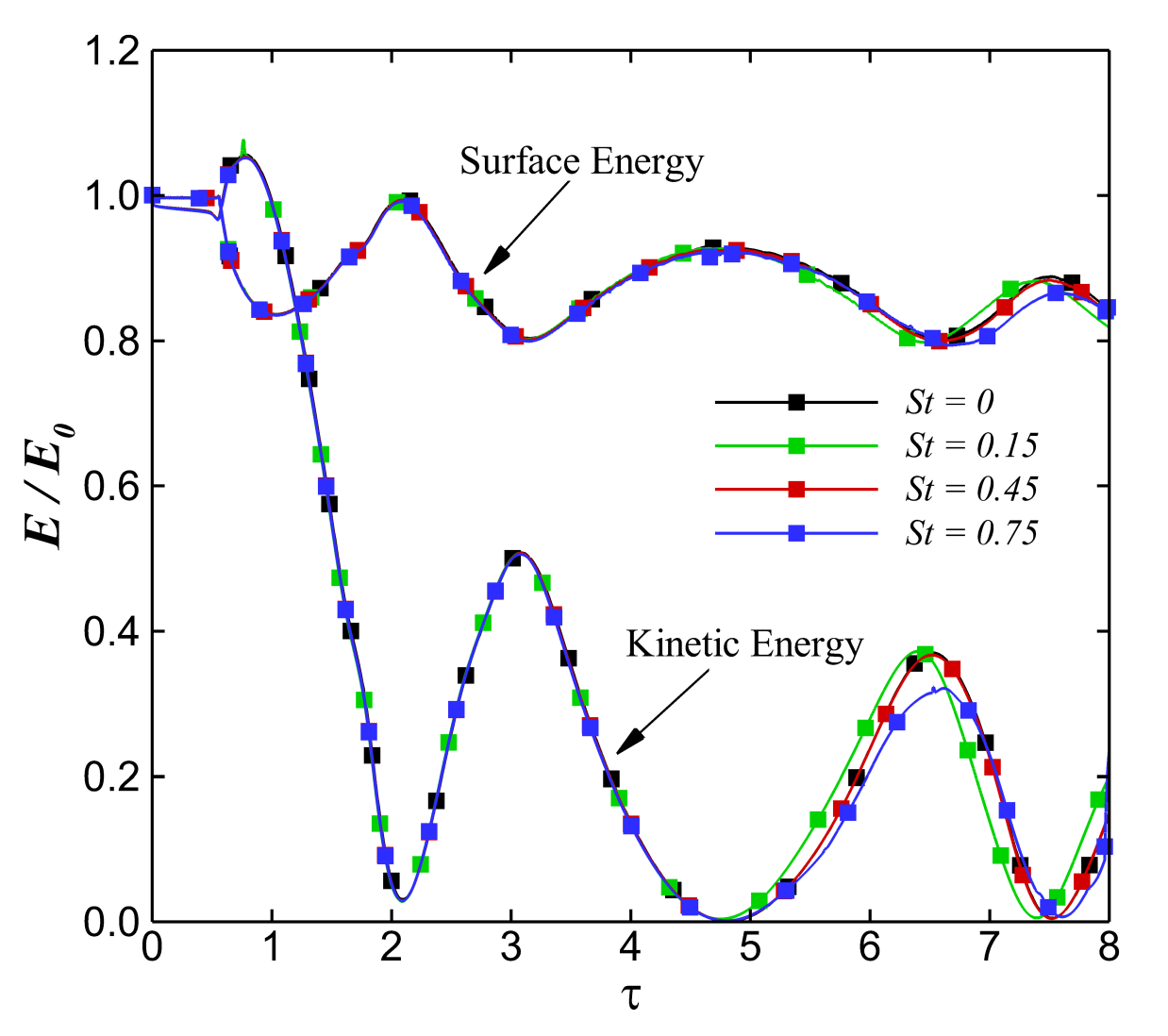}
\caption{Temporal evolution of normalized kinetic and surface energy of the drops in coalescence paradigm for different values of Stefan numbers. The rest of the parameters are $R_1=R_2=169.7$ $\mu$m, $We=13.63$, and $Oh=0.0354$.}
\label{fig:fig16}
\end{figure}

The collision of two identical tetradecane drops ($R_1=R_2=169.7$ $\mu$m) in the air is simulated in an axisymmetric domain (Fig. \ref{fig:fig1}). The boundary conditions for this case remain the same as in the previous case. The drops are given an initial relative velocity of $U=1.192$ m/s. The value of the dimensionless numbers are $We=13.63$ and $Oh=0.0354$. The evolution of the shape of the conglomerate without evaporation is shown in Fig. \ref{fig:fig13}. The interface ruptures as the drops come close to each other, and liquid from the two drops flows to form a single liquid volume. The liquid packet elongates in the radial direction due to the axial momentum of the drop fluid, creating the shape of a disk. After maximum compression in the axial direction, the liquid starts flowing towards the axis, and the elongation of the liquid along the axial direction takes place. This elongation and compression in axial and radial directions keep happening until the liquid mass loses all of its kinetic energy to viscous dissipation. In Fig. \ref{fig:fig14}, we present the contours of the temperature field and vapor mass fraction field along with the zero level-set for the case of $St=0.3$. First column shows the evolution of drop shape, temperature fields and vapor mass fraction fields are plotted in column 2 and 3 respectively. The evolution of total liquid volumes in the domain for different levels of superheat are compared in Fig. \ref{fig:fig15}. The liquid mass remains constant in the isothermal collision of the drops ($St=0$). As expected, the volume of the liquid decreases more with increasing the level of superheat, i.e. more evaporation. In Fig. \ref{fig:fig16}, we have shown the variation of the surface energy and the kinetic energy of the drops for different levels of evaporation. The coalescence of two drops is a process of competition between the surface and kinetic energies. As soon as the interface ruptures, the total surface energy takes a dip as the surface area of the drops decreases. The loss in surface energy increases the kinetic energy of the liquid, which in turn is responsible for the elongation of the liquid mass in the radial direction and its compression in the axial direction. It is evident from Fig. \ref{fig:fig16} that both the surface and kinetic energy keep oscillating with decreasing amplitude due to elongation and compression of the liquid mass. The oscillations are damped out, and the liquid mass takes a spherical shape eventually, after losing its kinetic energy. There seems to be almost no effect of evaporation on energy variations as the depletion in the mass of the liquid is very small. However, for the case of maximum evaporation ($St=0.75$), a slight decrease in the kinetic and surface energy of the liquid is observed as time progresses with the apparent occurrence of mass depletion.

\subsection{Reflexive separation with evaporation}\label{subsec:Ref_separation}

\begin{figure}[ht]
\centering
\includegraphics[width=0.95\textwidth]{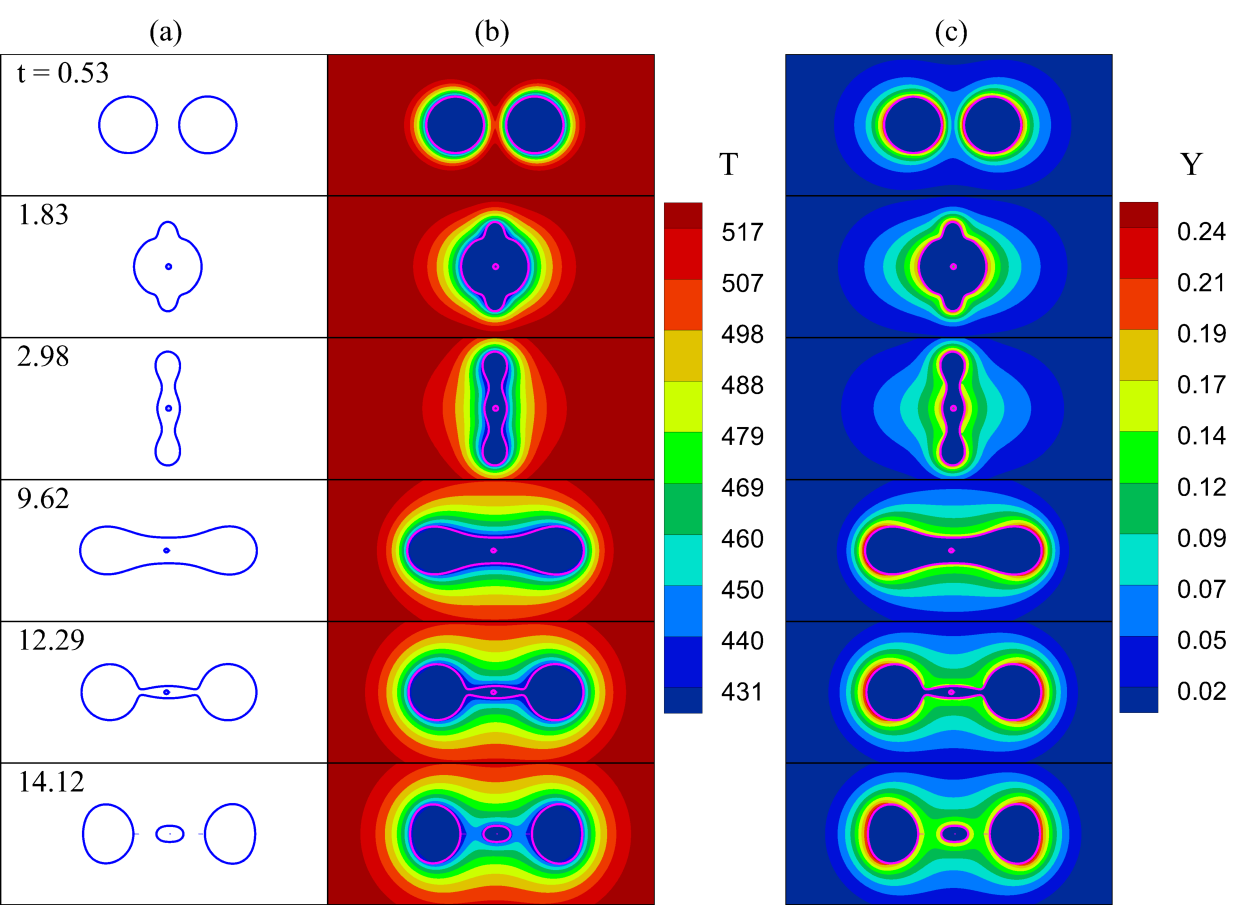}
\caption{Collision dynamics of two evaporating drops undergoing reflexive separation. Contours of the (a) interfaces of the drops characterised by the zero level-set function, (b) temperature field ($T$), and (c) vapor mass fraction ($Y$). The parameters used in the simulations are $R_1=R_2=150$ $\mu$m, $We=45.92$, $Oh=0.0376$, and $St=0.1$.}
\label{fig:fig17}
\end{figure}

\begin{figure}[ht]
\centering
\includegraphics[width=0.7\textwidth]{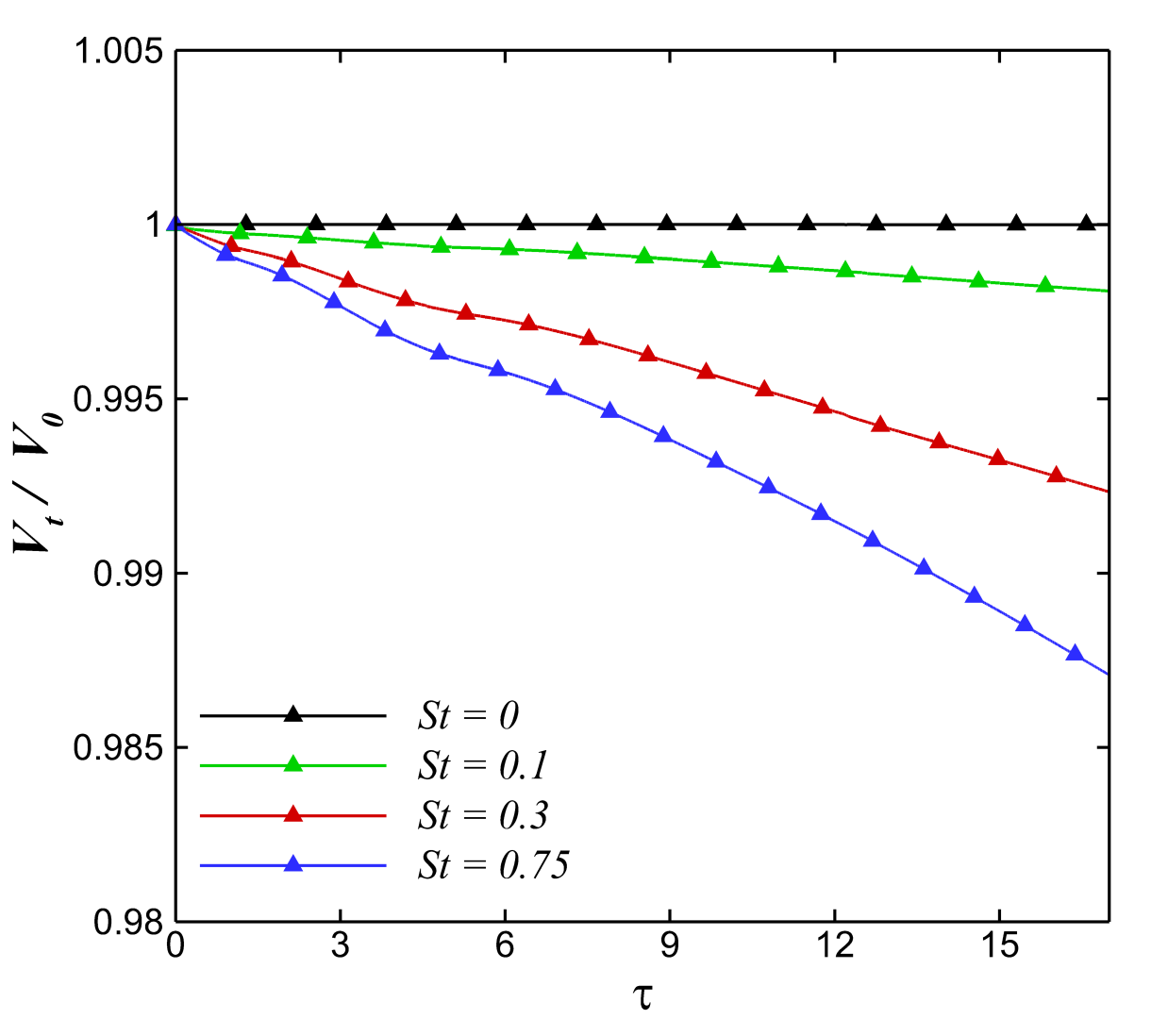}
\caption{Temporal evolution of the total liquid volume with the initial liquid volume of the drops for different values of Stefan numbers. The rest of the parameters are $R_1=R_2=150$ $\mu$m, $We=45.92$, and $Oh=0.0376$.}
\label{fig:fig18}
\end{figure}

Finally, we consider the case of two tetradecane drops colliding in hot air and undergoing reflexive separation. The pertinent governing input parameters are kept the same as those used in the case of Fig. \ref{fig:fig3}. The evaporation is activated by the solution of energy and vapor mass fraction equations together with the solution of the Navier-Stokes equations. Different levels of superheat of the ambient temperature are applied to study different evaporation rates from the drops. The pertinent dimensionless parameters used for the simulations are $R_1=R_2=150$ $\mu$m, $We=45.92$, and  $Oh=0.0376$. 

\begin{figure}[ht]
\centering
\includegraphics[width=0.7\textwidth]{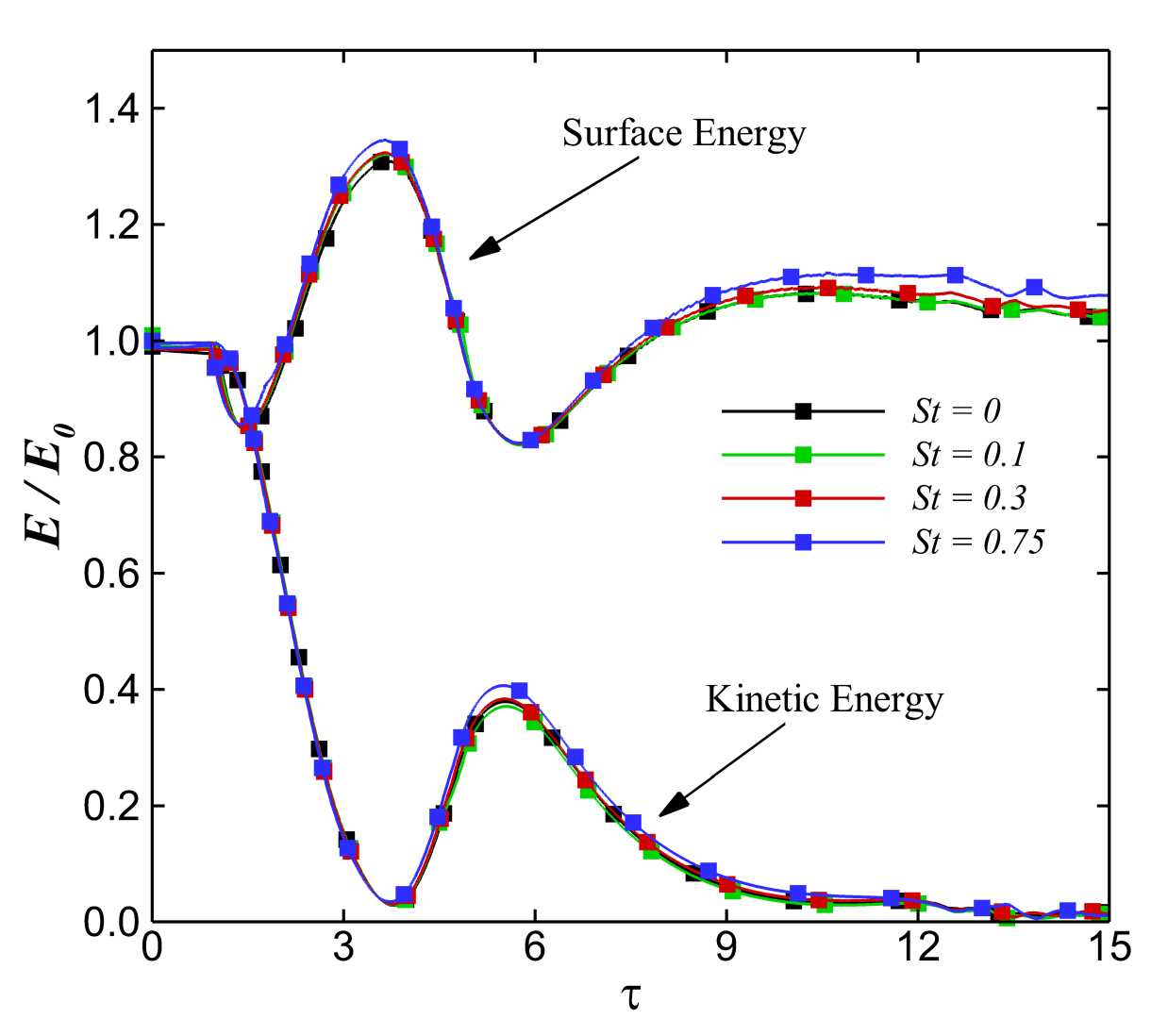}
\caption{Temporal evolution of normalized kinetic and surface energy of the drops in reflexive separation paradigm for different values of Stefan numbers. The rest of the parameters are $R_1=R_2=150$ $\mu$m, $We=45.92$, and $Oh=0.0376$.}
\label{fig:fig19}
\end{figure}

The drop-shape evolution and temperature and vapor mass fraction field contours are shown in Fig. \ref{fig:fig17} for $St=0.1$. The first column shows the contours of the interface of the drops (zero level-set function), temperature field $(T)$, and vapor mass fraction field $(Y)$ are shown in columns 2 and 3, respectively. Since, for this case, the Weber number is high enough, the drops have more inertial force. The drops coalesce with each other after the impact, and the liquid is compressed, forming a very thin disk. The liquid then starts elongating along the axis of symmetry, and the liquid volume is divided into two parts, joined by a ligament connecting the two liquid masses at their ends. Due to high inertia, the liquid masses pull themselves stretching the ligament. The neck starts to form at the two ends of the ligament due to the capillary forces (see panel 5 of Fig. \ref{fig:fig17} (a)). The pinch-off takes place at these two extreme ends of the ligament, known as the capillary break up \citep{hiranya2023}, giving rise to the formation of a small drop at the centre, which is known as the satellite drop. The evolution of total liquid volume with dimensionless time for different levels of superheat is plotted in Fig. \ref{fig:fig18}. As expected, the volume of liquid decreases more with increasing the superheat of the surrounding fluid. Figure \ref{fig:fig19} demonstrates the variation of the normalized kinetic and surface energy of the drops. In this process, the competition between kinetic and surface energy takes place. As the drops merge with each other, the surface energy takes a dip due to the loss in surface area, as in the case of coalescing drops. The liquid mass starts to elongate along the radial direction forming a thin disc. During this process, the surface energy starts increasing owing to the increase in surface area, thus reducing the kinetic energy. Due to the influence of capillary forces, the liquid mass is pulled back towards the axis of symmetry, and finally, it starts elongating in the axial direction. The conversion of surface energy into kinetic energy happens during this movement of the liquid mass. The kinetic energy of the liquid keeps on decreasing as the mass pulls apart. The pinch-off takes place, forming a satellite drop. The kinetic energy of the liquid decreases further until the liquid masses come to rest. The oscillations of the kinetic energy do not take place as was observed in the case of coalescence.

\section{Conclusions}\label{sec:conclusion}

This study investigates the dynamics of head-on collisions between two drops in the presence of evaporation. Numerical simulations are employed to analyze various collision outcomes, including bouncing, coalescence, and reflexive separation. The validation process demonstrates good agreement, particularly for bouncing and reflexive separation collisions. The evaporation model accurately predicts the phase change from liquid to vapor. Simulations are conducted with different levels of superheat to examine the impact of varying evaporation rates. Contours of the temperature and vapor mass fraction fields are presented for all three collision types. The simulations reveal the entrapment of vapor in the case of bouncing separation with evaporation, preventing the drop interfaces from making contact. The vapor attempts to escape the entrapment with high velocity. Increasing evaporation rates result in a slower separation of the drops after collision. The difference between them is illustrated by plotting the normalized kinetic and surface energy values for different Stefan numbers. In the case of coalescence, the kinetic and surface energies of the drops exhibit oscillations until the kinetic energy is completely dissipated. However, no such oscillations occur in drops undergoing bouncing and reflexive separations. The drops do not fully recover their kinetic energy after the collision for bouncing separation due to the viscous effect. In the reflexive separation regime, the kinetic energy of the drops becomes zero after detachment.

\section*{Acknowledgements:} A.K.P. acknowledges the support from PMRF fellowship, Government of India and G.B. acknowledges his gratitude to J. C. Bose National Fellowship of SERB, Government of India (grant: JBR/2020/000042). K.C.S. thanks the Science and Engineering Research Board, India, for the financial support through grant CRG/2020/000507. The authors express their sincere appreciation to the high-performance computing facilities (HPC and Paramsanganak) of IIT Kanpur for the valuable support.


 \end{document}